\documentclass[12pt]{article}
\usepackage[tablesfirst,notablist,nomarkers]{endfloat}

\usepackage{ol2}
\usepackage{amsmath}

\begin{document}

\title{Time gating of heralded single photons for atomic memories}

\author{B. Melholt Nielsen,$^1$ J. S. Neergaard-Nielsen,$^1$ and E. S. Polzik$^{1,*}$}

\address{
$^1$Niels Bohr Institute, Danish National Research Foundation Center for Quantum Optics
\\ (QUANTOP), Blegdamsvej 17, DK-2100 Copenhagen, Denmark
\\
$^*$Corresponding author: polzik@nbi.dk
}

\begin{abstract}We demonstrate a method for time gating the standard heralded continuous-wave (cw) spontaneous parametric down-converted (SPDC) single photon source by using pulsed pumping of the optical parametric oscillator (OPO) below threshold. The narrow bandwidth, high purity, high spectral brightness and the pseudo-deterministic character make the source highly suitable for light-atom interfaces with atomic memories.\end{abstract}

\ocis{270.0270, 270.5290, 270.5530, 270.6570, 030.5260, 190.4970.}

\noindent In quantum communication and quantum information processing light is the natural agent for carrying information, whereas atomic systems are highly suitable for storing and processing information. An efficient exchange of quantum information between photonic carriers and atomic nodes becomes important since it would allow for the implementation of quantum information networks.

In a number of approaches to quantum memory \cite{hammerer-etal:2008}, such as electromagnetically induced transparency \cite{Eisaman:2005,Chaneliere:2005}, Faraday interaction with feedback \cite{julsgaard-etal:2004}, and teleportation \cite{sherson-etal:2006} the temporal mode of the quantum field of interest has to be matched well with that of a strong driving field to achieve a successful storage. An attractive light source for such atomic memories would be the one that generates pure non-classical quantum states in a deterministic manner.

Single photon sources based on single emitters in free space usually suffer from poor collection efficiency. This deficiency can be remedied by placing them inside optical cavities \cite{McKeever:2004,Pelton:2002,Legero:2004}. Atomic ensembles can serve as sources of single photons \cite{Eisaman:2005,Chaneliere:2005,Chou:2004}, however the efficiency to-date is not very high. The highest purity single photon sources up to date are based on heralded SPDC where one of the two downconverted photons is counted and heralds the presence of the other photon. The SPDC sources come in a pulsed free space version \cite{lvovsky-etal:prl:2001,zavatta-etal:pra:2004,pittman-etal:oc:2005,ourjoumtsev-etal:prl:2006} and in a cavity-enhanced cw setting based on an OPO \cite{neergaard-nielsen-etal:oe:2007}. The narrow bandwidth, perfect spatial mode and high spectral brightness of the cavity-enhanced setting makes it well suited for application with atomic memories. One major disadvantage of the cw OPO setting is the random character of the state generation.

In this Letter we demonstrate how to generate high purity time gated single photons using a pulsed OPO pump in the cw SPDC setup, thus combining the superior spectral and spatial properties of the OPO output with the enhanced regularity of the pulsed pump approach. The OPO pump pulses impose a gating condition on the single photon production whereby the creation of photons inside the OPO cavity is only possible within the time window of the pump pulses. The particular time gates in which the SPDC photon actually appears are heralded by the idler photon.  One can thus synchronize the source with the driving field of an atomic memory protocol, and then only keep the result in the memory if the heralding photon has been counted. In our experiment we perform a tomographic reconstruction of the single photon using a pulsed temporal mode function, a cw local oscillator (LO) and a homodyne detector (HD).

\begin{figure}[htb]
\centerline{\includegraphics[width=8.3cm]{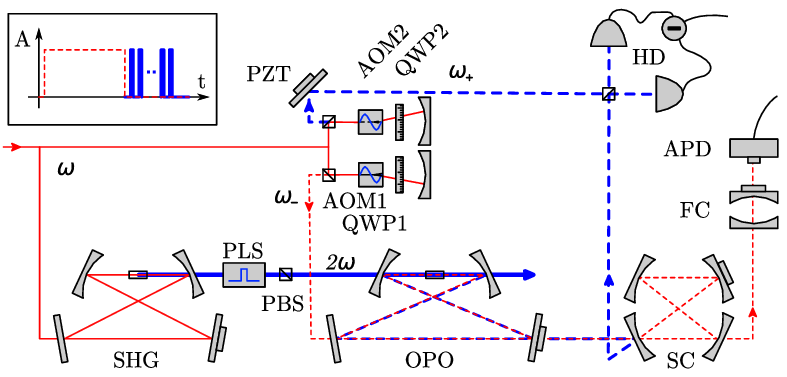}}
\caption{(Color online) The experimental setup. Inset shows the experimental cycle of locking and pulsing.}
\label{fig01}
\end{figure}

A master beam is derived from a cw Ti:Sapphire laser running at the Cesium D2 line $\omega$ as shown in Fig. \ref{fig01}. This beam is split to pump the second harmonic generator (SHG) and to serve for the production of a blue-shifted beam at $\omega_{+}$ and a red-shifted beam at $\omega_{-}$. Both are shifted by one OPO free spectral range $\omega_{FSR}$ to match the first non-degenerate components of the OPO output at $\omega_{\pm} = \omega \pm \omega_{FSR}$. The red-shifted  beam is used as a locking beam for the OPO, the split-off cavity (SC) and the linear filter cavities (FC), while the blue-shifted beam is used as a LO for the homodyne detector. The OPO pump is pulsed using a pulse-shaper (PLS) and the generated photon pairs are split on the SC. The photon created at $\omega_{-}$ is transmitted and spectrally filtered using the FC before impinging on the avalanche photo-diode (APD) whereby heralding the presence of its twin photon created at $\omega_{+}$. The twin photon is reflected off the SC and sent onto the HD for tomography. More details can be found in \cite{neergaard-nielsen-etal:oe:2007}.

The PLS is a Pockels cell driver (BME Bergmann) with a KD*P Pockels cell and a polarizing beam splitter (PBS). The maximum repetition rate is 200kHz and the optical rise time is 5ns. The pulse extinction ratio is $1.5 \cdot 10^{-2}$ which is mainly due to the poor quality of the PBS. The pulse length is controlled by electronically delaying the signal that turns the pulse off.

Within one experimental cycle we first dither the length of the OPO, SC and FC around their resonances using the red-shifted beam as a locking beam and monitoring the APD signal to tune all cavities to resonance.  When the cavities are on resonance we turn off the red-shifted beam with the acousto-optical modulator (AOM1) and let them stand passively while pulsing the OPO pump with a 50kHz repetition rate during the measurement (see inset in Fig. \ref{fig01}). We measure for 200ms and lock for 800ms. The OPO, SC and FC are able of passively staying near resonance for up to 10s.

\begin{figure}[htb]
\centerline{\includegraphics[width=8.3cm]{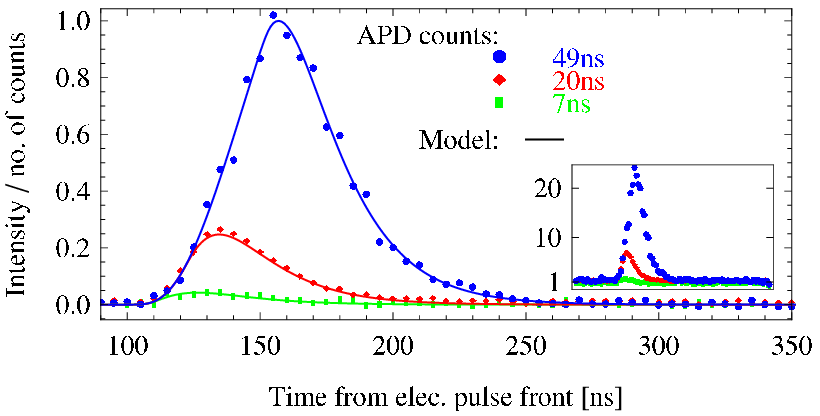}}
\caption{(Color online) The photon counting distribution. Comparing model and data. Solid lines are model output through OPO, SC and FC. Points are the rescaled APD count data. Inset shows the 500ns long raw distribution normalized to the background level.}
\label{fig02}
\end{figure}

We first measure the photon counting distribution of the field arriving at the APD. This is done using a digital oscilloscope (Lecroy Wavepro 7100) to record the time delay of the APD counts occurring within 500ns from the front of the electronic pulse driving the PLS. Histograms of measured APD click delay times for three different optical pulse lengths are shown in the inset of Fig. \ref{fig02}. The raw count rates are 4.4$s^{-1}$, 5.7$s^{-1}$ and 27$s^{-1}$ for pulses of length 7ns, 20ns and 49ns. Without pulsing the corresponding rate is 275$s^{-1}$.  The background count rate is 3.8$s^{-1}$ and is mainly due to the poor extinction ratio of the PLS since the APD dark count is 0.4$s^{-1}$. It should be possible to improve the count rate by increasing the repetition rate and measurement time window.

To model the photon counting distribution we consider propagation of the optical pulse through the OPO, SC and FC. The pulse as a square with the rise time of 5ns, and the field transmission through a cavity corresponds to applying a Lorentzian frequency filter,
\begin{align}
	F\left( \omega \right) & \propto \frac{ \kappa }{ \kappa^{2}+\omega^{2} }, \label{eq01}
\end{align}
\noindent where $\kappa$ is half the cavity bandwidth. The output field is found by convolution,
\begin{align}
	z\left(t\right) & \propto  \int_{-\infty}^{t} z_{p}\left(\tau\right) F\left(t-\tau\right) d\tau, \label{eq02}
\end{align}
where $z_{p}\left(t\right)$ is the temporal profile of the pulse and $F\left(t\right)$ is the time-domain response of the product of the frequency filters of the OPO, $\gamma/2\pi = 4.4$MHz, and the most narrow filter cavity, $\kappa/2\pi = 12$MHz. This rough model is compared to the APD count data in Fig. \ref{fig02}. The data has been scaled vertically by the same factor to fit the model and the background level has been removed.

\begin{figure}[htb]
\centerline{\includegraphics[width=8.3cm]{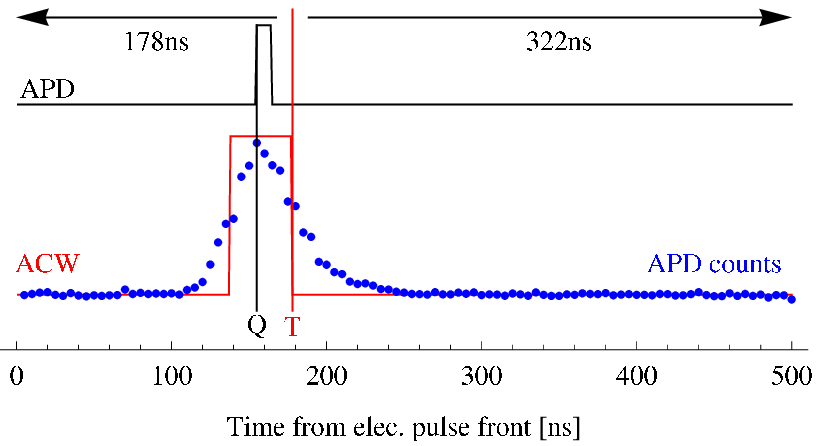}}
\caption{(Color online) Homodyne data acquisition. The APD click serves as a qualifier (Q). The trailing edge of the acceptance window (ACW) serves as a trigger (T) if and only if it occurs within 40ns after the qualifier. An APD count distribution is shown for comparison.}
\label{fig03}
\end{figure}
The homodyne measurement is done using the oscilloscope to record homodyne data conditioned on an APD click occurring within a 40ns long acceptance window as shown in Fig. \ref{fig03}. The acceptance window is centered on the time of the peak of the photon count distribution. On every click we sample a 500ns long time window from the front edge of the electronic pulse. The sample rate is $10^{9}s^{-1}$ and we collect data from a few $10^{4}$ clicks.

\begin{figure}[htb]
\centerline{\includegraphics[width=8.3cm]{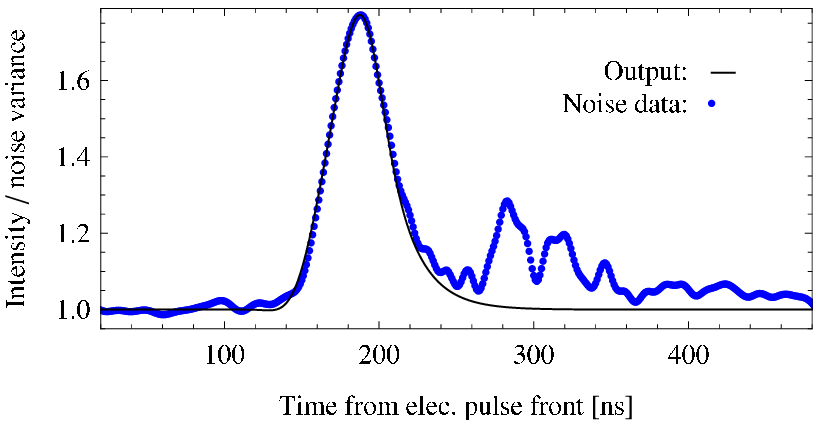}}
\caption{(Color online) The noise variance of 13,000 time windows for a 49ns long pulse. Solid line is a model pulse filtered by the OPO and low-pass filtered for comparison.}
\label{fig04}
\end{figure}

We estimate the temporal mode function for post processing the homodyne data by looking at the variance of the individual points between successive time windows - see Fig. \ref{fig04} for the case of a 49ns long pulse. The noise variance has been normalized to the vacuum level and filtered using a 25MHz low-pass filter to remove some high-frequency electronic noise due to the experimental hardware. The ripples after the peak in Fig. \ref{fig04} are electronic noise from the APD and the additional time delay compared to that of Fig. \ref{fig02} is due to the difference in the optical and electronic paths.

In \cite{nielsen-molmer:pra:2007:3} the task of finding the optimal mode function for the pulsed pump setting is addressed. There it is shown that in the limit where the acceptance window is much longer than the round-trip time of the OPO a good approximation for the optimal mode function is the temporal profile of the pulse convoluted by the OPO filter. Since our round-trip time is 2.7ns we are roughly in that limit with a 40ns long acceptance window. A comparison of this optimal mode function with the experimental noise variance is given in Fig. \ref{fig04}.

\begin{figure}[htb]
\centerline{\includegraphics[width=8.3cm]{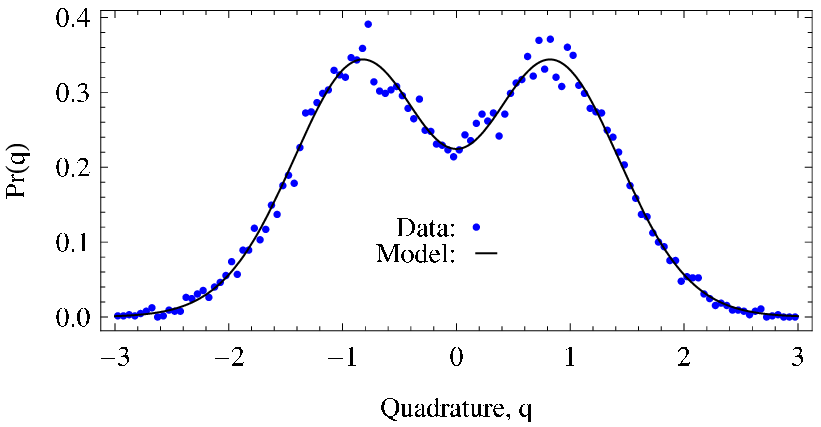}}
\caption{(Color online) Marginal distribution of 13,000 quadrature points. The points are the phase-averaged marginal distribution and the solid line is the model fit.}
\label{fig05}
\end{figure}

The post processing is done by multiplying the data from each time window by the mode function and integrating to obtain one quadrature point. These quadrature points are then normalized to a vacuum level. Assuming the generated state to be phase invariant we use all the quadrature points to obtain the phase-averaged marginal distribution shown in Fig. \ref{fig05}.

We take the generated state to be a sum of Fock states $n = 0,1,\dots,6$ and fit this model to the marginal distribution. The best fit is shown in Fig. \ref{fig05}. The first three diagonal elements of the density matrix are $\rho_{00} = 0.392$, $\rho_{11} = 0.595$ and $\rho_{22} = 0.010$ while the rest are of order $10^{-3}$ or less. The center value of the Wigner function is $W\left(0,0\right) =  \left(\rho_{00}-\rho_{11}+\rho_{22}-\cdots\right)/\pi$ and using the above diagonal elements it evaluates to $W\left(0,0\right) = -0.061$. It should be noted that these numbers have not been corrected for any experimental inefficiencies.

A similar homodyne measurement using a 20ns long pump pulse resulted in a somewhat worse generated single photon state, although it should have been better because of a better defined temporal mode according to \cite{nielsen-molmer:pra:2007:3}. The first two diagonal elements of the phase-averaged density matrix of 4,200 quadrature points are $\rho_{00} = 0.542$ and $\rho_{11} = 0.458$, while the rest are vanishingly small. The center value of the Wigner function for these diagonal elements is $W\left(0,0\right) = 0.027$. We believe the reason for this is the background level visible in the inset of Fig. \ref{fig02}. The ``signal-to-noise'' ratio is worse for the 20ns pump than for the 49ns pump and this implies that we sample more vacuum and less single photons.

The source presented here is of high purity comparable to the two SPDC-based sources with the highest purity reported to-date in \cite{pittman-etal:oc:2005} (pulsed, purity 80\%, rate 1,000$s^{-1}$, bandwidth 5THz) and in our earlier work \cite{neergaard-nielsen-etal:oe:2007} (cw, purity 70\%, rate 13,000$s^{-1}$, bandwidth 8MHz). Its spectral brightness is $10^{4}$ times higher than that of the free-space source \cite{pittman-etal:oc:2005}. The spectral brightness reported here is 500 times lower than that of the heralded cw source \cite{neergaard-nielsen-etal:oe:2007} which is the price we pay for the gated character of the present source.

In summary we have demonstrated the improved regularity of the heralded cw SPDC single photon source by time gating the photon pair production using the pulsed pumping of the OPO. The time gating makes it possible to synchronize the single photon with atomic memory protocols by choosing the driving field pulse for the memory interaction according to the suitable temporal mode function. The generated single photons have a bandwidth of a few MHz suitable for atomic memories, a well-defined spatial mode, high purity and high spectral brightness.

This research was funded by European grants QAP, COMPAS, and EMALI.

\bibliographystyle{ol}

\end{document}